# THE IMPACT OF MEDICAID COVERAGE ON MENTAL HEALTH, WHY INSURANCE MAKES PEOPLE HAPPIER IN OHIE: BY SPENDING LESS OR BY SPENDING MORE?

Yangyang Li*

*Department of Electrical Engineering and Computer Science and the Department of Economics, Massachusetts Institute of Technology, United States*

**Abstract:** The Oregon Health Insurance Experiment (OHIE) offers a unique opportunity to examine the causal relationship between Medicaid coverage and happiness among low-income adults, using an experimental design. This study leverages data from comprehensive surveys conducted at 0 and 12 months post-treatment. Previous studies based on OHIE have shown that individuals receiving Medicaid exhibited a significant improvement in mental health compared to those who did not receive coverage. The primary objective is to explore how Medicaid coverage impacts happiness, specifically analyzing in which direction do variations in healthcare spending significantly improve mental health: higher spending or lower spending after Medicaid. Utilizing instrumental variable (IV) regression, I conducted six separate regressions across subgroups categorized by expenditure levels and happiness ratings, and the results reveal distinct patterns. Enrolling in OHP has significantly decreased the probability of experiencing unhappiness, regardless of whether individuals had high or low medical spending. Additionally, it decreased the probability of being pretty happy and having high medical expenses, while increasing the probability among those with lower expenses. Concerning the probability of being very happy, the OHP only had a positive effect on being very happy and spending less, and its effect on those with high expenses was insignificant. These findings align with the benefit of Medicaid: alleviating financial burden, contributing to the well-being of distinct subgroups.

## Introduction

The advent of the Oregon Health Insurance Experiment (OHIE) in 2008 offered an unprecedented opportunity to rigorously evaluate the causal effects of Medicaid coverage on a range of outcomes through a randomized controlled design (Finkelstein et al., 2012). As the state of Oregon opened its Medicaid program to a limited number of low-income adults via a lottery system, a natural experiment unfolded from the 90,000 individuals who signed up, allowing for an objective analysis that sidesteps the perennial challenges of unobserved differences that often confound such research (Baicker & Finkelstein, 2011).

This paper contributes to the body of evidence by examining the nuanced effects of Medicaid coverage on mental health and happiness within the OHIE framework. Specifically, it investigates whether increased happiness among Medicaid recipients is attributable to greater healthcare spending, which may imply improved access to necessary services, or to reduced financial strain due to lower out-of-pocket expenditures.

The Oregon Medicaid lottery, by effectively randomizing access to public insurance, mitigates the selection bias inherent in previous observational studies. The resulting analysis leverages both administrative and survey data to glean insights into the complex dynamics at play between health insurance, healthcare utilization, and subjective well-being.

*Corresponding Author's Email:*annieliy@mit.edu

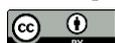

Initial findings from the OHIE indicate that Medicaid coverage has led to statistically significant higher healthcare utilization, reduced out-of-pocket expenses, and enhanced self-reported health among the lottery-selected group as compared to the control group without such coverage (Taubman et al., 2014). This research extends these findings by disaggregating the effects of Medicaid coverage on happiness. The results in this research show an overall trend of a decrease in the probability of being pretty happy and having high medical expenses, and an increase in the probability among those with lower expenses. Concerning the probability of being very happy, the OHP only had a positive effect on being very happy and spending less, and its effect on those with high expenses was insignificant. These results align with the advantages of Medicaid, showing how it helps by reducing financial stress and improving the overall health of specific groups within the population

Background

The Oregon Health Insurance Experiment (OHIE) has served as a seminal source for understanding the effects of Medicaid expansion. Taubman et al. (2014) provided crucial insights, reporting that Medicaid significantly increased emergency department visits by 0.41 visits per person, contradicting the hypothesis that Medicaid expansion would decrease costly emergency department usage by improving access to primary care and overall health (Taubman et al., 2014). Finkelstein et al. (2012) expanded on these findings by demonstrating that, in the first year after Medicaid expansion, there was a significant increase in healthcare utilization including primary and preventive care, alongside reductions in financial strain due to medical expenses. Their work underscored the complexity of healthcare behaviors and hinted at potential long-term benefits not immediately evident in terms of cost reductions or health outcomes (Finkelstein et al., 2012). Kowalski (2016) advanced the analytical approach by applying Marginal Treatment Effect (MTE) methods to dissect heterogeneity within the OHIE data. Her work revealed that the treatment effect of insurance on emergency room utilization varied significantly across different subgroups, delineating a more nuanced understanding of how Medicaid impacts different population segments (Kowalski, 2016). Kowalski (2018) further contributed by comparing results from Oregon with the Massachusetts health reform, aiming to reconcile why similar expansions led to divergent outcomes in emergency department utilization. By leveraging the MTE framework, she suggested that initial health status and prior healthcare utilization patterns could explain these variations, proposing that healthier new enrollees in Massachusetts might reduce emergency usage, unlike in Oregon (Kowalski, 2018).

The body of research emanating from the OHIE underscores the intricate dynamics of healthcare policy's impact on low-income populations (Baicker et al., 2014). These studies collectively emphasize the need for nuanced policy instruments that consider initial health conditions, existing healthcare infrastructure, and localized healthcare behaviors. Further research should continue to leverage randomized designs where feasible, alongside sophisticated econometric models to untangle the causal impacts of health policy changes (Kaczynski & Solnica, 2012).

**Data**

Data source

The dataset I use is the Oregon Health Insurance Experiment (OHIE) from the National Bureau of Economic Research (NBER) (National Bureau of Economic Research, n.d.). The OHIE dataset covers a period from March to September 2008, when the Medicaid lottery was conducted, with follow-up data collected primarily at one and two years post-lottery, extending the coverage of the dataset through at least 2010, which allows the analysis of short-term impacts of Medicaid coverage on the participants (Finkelstein et al., 2012).

The unit of observation for the data in the OHIE is at the individual level (Finkelstein et al., 2012). The lotterfy system selected individuals to have the opportunity to apply for Medicaid, and the results were analyzed based on individual health outcomes, financial hardship, and other factors (Baicker et al., 2013). Additionally, since the opportunity to apply for coverage could extend to other family members within the household, the analysis also took into account household-level effects (Finkelstein et al., 2012). In 2008, 89,824 individuals entered the lottery, with 35,169 individuals, representing 29,664 households, selected for the chance to apply for coverage (Taubman et al., 2014).

The OHIE dataset is a longitudinal panel dataset that captures cross-sectional data at multiple points in time (Hattab et al., 2024). It is primarily a cross-sectional dataset with elements of longitudinal tracking in two periods. It is not a traditional time series dataset, because it does not track variables continuously over time; instead, it captures data at specific points following the implementation of the Medicaid lottery. However, it does follow the same individuals over a period, thereby incorporating some longitudinal aspects.

The data I use are composed of the descriptive data and the survey data collected after the lottery. Descriptive data were collected when individuals entered the lottery and then again during follow-up periods to assess outcomes after the lottery. The follow-up data included several waves, with each capturing data post-lottery, half-year after lottery and one year after lottery, thus forming repeated cross-sections of data for the individuals and households involved (National Bureau of Economic Research, n.d.). Applicants for the lottery provided characteristic information, while follow-up surveys collected healthcare utilization data, including emergency department visits, hospital admissions, prescription drug use, and financial measures such as credit scores and debts.1

Summary Statistics

This study utilizes data from a series of surveys conducted at different intervals from 0 months and 12 months after participants received treatment. Notably, the dataset corresponding to the 6-month interval was excluded from analysis due to significant data incompleteness and inadequate temporal coverage post-treatment initiation, precluding a comprehensive assessment of Medicaid's effect during this period. Our aim is to understand how having Medicaid coverage influences happiness, focusing on whether decreased healthcare spending or increased healthcare spending contributes more significantly to this outcome.

This study utilized a dataset I merged from three datasets of OHIE Public Use Data, as listed below:

Descriptive Data (named as `oregonhie_descriptive_vars`): Contains descriptive variables about the participants, such as personal and household IDs and lottery selection status.

First and Final Wave Survey Data (named as `oregonhie_survey0m_vars` and `oregonhie_survey12m_vars`): Include variables from surveys conducted immediately after receiving treatment and 12 months later, respectively.

These datasets provide a comprehensive view of the participants' demographic information, Medicaid coverage status, healthcare spending, and reported happiness over time, etc.

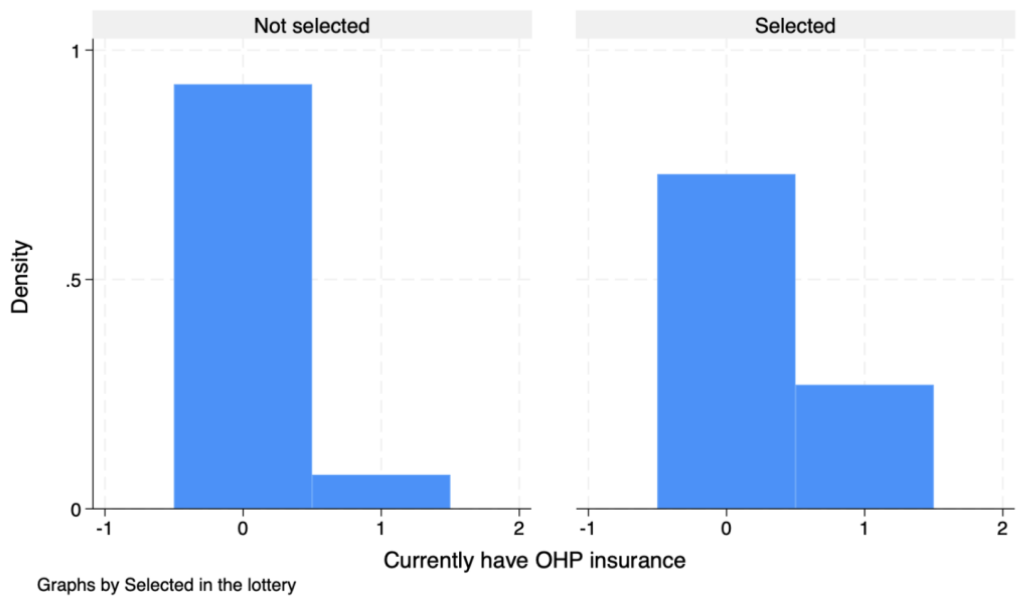

Figure 1: Treatment Intensity

Figure 1 illustrates the distribution of current Oregon Health Plan (OHP) insurance coverage among individuals selected and not selected in the lottery. The left panel shows that individuals not selected in the lottery exhibit a near-zero density of having OHP insurance, as expected. Conversely, the right panel indicates that those selected in the lottery have a bimodal distribution, with significant proportions both possessing and not possessing OHP insurance.

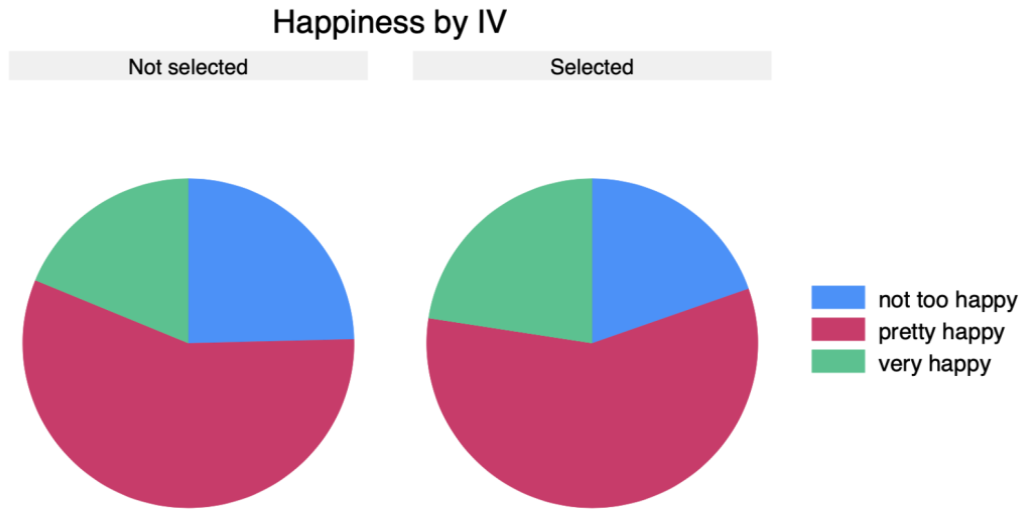

Figure 2: Happiness by IV

Figure 2, titled "Happiness by IV," compares the self-reported happiness levels of individuals who were not selected versus those who were selected in the lottery to receive free insurance. Both pie charts are segmented into three categories of happiness: "not too happy," "pretty happy," and "very happy." It appears that the proportion of individuals reporting higher happiness ("very happy") is slightly larger in the group selected to receive insurance compared to those not selected. This visual suggests a potential positive association between being selected for Medicaid coverage and self-reported happiness. Further analysis in this study will be done to establish causality and quantify the effect size.

Data cleaning and Variable Definitions

To prepare the data for analysis, several steps were taken:

Participants Selection: Only those who participated in both the initial and 12-month follow-up surveys were retained, resulting in the deletion of 16,517 records.

Missing Data Handling: Missing answers for gender were addressed by making one adjustment. A significant cleaning step involved rectifying 406 cases with missing treatment variables.

Outlier Deletion: One identification and deletion of outliers that could potentially skew the results. Specifically, an extreme value of expenditure, recorded at 2.20e+7, was identified as an outlier and dropped, resulting in the deletion of one observation and 58,404 non missing values remaining.

Treatment Variable Correction: I utilized enrollment in OHP in the first wave survey data as the treatment variable. To ensure the accuracy of the treatment variable, a correction was applied based on additional data fields: for individuals with data indicating confirmed acceptance into OHP by both preliminary and final approval processes, the treatment status was uniformly updated to one. This correction was necessary to accurately reflect the enrollment status in the OHP and reduce missing data problems, as these conditions conclusively demonstrate that the participants were accepted into the plan.

Variable Generation: Age variable was generated from the year of birth.

Table 1: Summary Statistics

| Variable | N | Mean | Standard Deviation | Min | Max |
|---|---|---|---|---|---|
| Winning Lottery (IV) | 58404 | 0.5066 | 0.5000 | 0 | 1 |
| OHP (treatment) | 23139 | 0.1734 | 0.3786 | 0 | 1 |
| Happiness (outcome) | 23449 | 1.7365 | 0.6585 | 1 | 3 |
| # HH Members | 58404 | 1.2927 | 0.4608 | 1 | 3 |
| Expenditure | 22765 | 736.3079 | 55076.7200 | 0 | 8301400 |
| Age | 58404 | 39.8860 | 12.1225 | 20 | 63 |
| Female | 58404 | 0.5464 | 0.4978 | 0 | 1 |
| Previous Expenditure | 24028 | 583613.4 | 8.39e+07 | 0 | 1.30e+10 |

After major data cleaning process, key variables were generated and renamed for clarity:

- `IV (Z_i): Winning the Lottery`: Named Winning Lottery in Table 1, indicates lottery selection, a binary variable identifying if participants were selected (1) or not (0). 58,404 non missing values, binary variable. Originally named treatment, renamed to LotterSelected.

- `Treatment (D_i): Enrollment in the OHP`: The treatment variable, designated as D, was initially labeled as ins_ohp_0m in the dataset and was renamed to OHP in Table 1, identifying whether participants currently have coverage under the Oregon Health Plan (OHP). This variable is a binary indicator, where a value of 1 signifies that the participant has OHP coverage, and a value of 0 indicates the absence of such coverage. Data for the treatment variable were collected from 23,139 non-missing entries 12 months following the lottery selection. This rigorous adjustment ensures that the treatment variable reliably represents the actual health coverage status of the participants, thereby facilitating a more precise analysis of the impact of OHP coverage on their reported happiness outcomes.

- `Outcome ($Y_i$): Level of Happiness`: Named Happiness in Table 1, is a dependent variable. Represents the reported overall happiness, not too happy (1), pretty happy (2), and very happy (3). Collected 12 months after treatment with 23,449 non missing values. Originally named happiness_12m.

- `Control ($X_i$)`: The first control variable is the number of people in a household on the lottery list with 58,404 non missing values. Originally named numhh_list, renamed to # HH Members in Table 1. Currently, other control variables of my selection include age, gender, and previous expenditure, named Age, Female, and Previous Expenditure in Table 1, all have 58,404 non missing values.

- `Expenditure ($E_i$)`: Total out-of-pocket healthcare spending over the last 12 months with 22,765 non missing values, in dollars. Originally named cost_tot_oop_12m, renamed to Expenditure in Table 1.

- `median (median)`: median of all $E_i$.

- `Above Median ($AM_i$)`: dummy variable, $AM_i$ equals to 1 if above median of all $E_i$, 0 otherwise.

**Empirical Methodology**

Identification Strategy

The instrument utilized is the lottery selection ($Z_i$), which is assumed to be random and thus uncorrelated with unobservable determinants of the outcome variable (happiness). The instrument's validity is based on its correlation with the treatment variable ($D_i$), which represents a self-selection process in which a lottery-selected individual can choose to enroll in OHP/Medicaid coverage or not, while being uncorrelated with the error term in the outcome equation.

The empirical strategy for assessing the impact of Medicaid coverage on happiness involves a two-stage least squares (2SLS) regression, utilizing an instrumental variable (IV) approach. This method addresses potential endogeneity in the treatment assignment (Medicaid coverage).

The 6 Regressions Model

I perform subgroup analyses to determine if the impact of Medicaid coverage on happiness differs by levels of medical expenditure ($E_i$). This involves running separate regressions for subgroups where expenditures are $AM_i = 0$ (below the median) and $AM_i = 1$ (above the median), happiness $Y_i = 1$ (not too happy), $Y_i = 2$ (pretty happy), and $Y_i = 3$ (very happy). I first create six dummy variables and rename them $\tilde{Y}_{(j,i)}$, defined as:

- $\tilde{Y}_{(1,i)} = 1(Y_i=1, AM_i=1)$

- $\tilde{Y}_{(2,i)} = 1(Y_i=1, AM_i=0)$

- $\tilde{Y}_{(3,i)} = 1(Y_i=2, AM_i=1)$

- $\tilde{Y}_{(4,i)} = 1(Y_i=2, AM_i=0)$

- $\tilde{Y}_{(5,i)} = 1(Y_i=3, AM_i=1)$

- $\tilde{Y}_{(6,i)} = 1(Y_i=3, AM_i=0)$

The regression model can be written as:

$\tilde{Y}_j = \beta_{(j,0)} + \beta_{(j,1)} D + X^{\wedge'} \beta_{(j,X)} + \epsilon, j=1,\cdots,6$   (Second Stage) $D = \gamma_0 + \gamma_1 Z + X^{\wedge'} [\![\gamma_X]\!]\_ + u$ (First Stage)

$\beta_{(j,1)}$ captures the effect of enrolling in OHP on the probabilities, and is the coefficient I want to identify and estimate. For example, when $j=1$, $\beta_{1,1}$ can be written as:

$\beta_{1,1} = E[1(Y_i=1, AM_i=1)|D_i=1] - E[1(Y_i=1, AM_i=1)|D_i=0]$

$= Pr[Y_i=1, AM_i=1|D_i=1] - Pr[Y_i=1, AM_i=1|D_i=0]$.

The transformation from expectation to probability is because the outcome is binary.

Assumptions

The following are assumptions for the 2SLS regression analysis:

(1) $\{Y_i, D_i, Z_i, X_i\}_{(i=1)}^{\wedge N}$ i.i.d.

(2) $E(\epsilon_i | Z_i) = 0, E(u_i | Z_i) = 0$

(3) $corr(D_i, Z_i) > 0$

(4) $var(\epsilon_i | Z_i) < \infty, var(u_i | Z_i) < \infty$

The interpretations of these assumptions are:

(1) The observations of the outcome variable ($Y_i$), the treatment variable ($D_i$), the instrumental variable ($Z_i$), and the control variables ($X_i$) are presumed to be independent and identically distributed (i.i.d.).

(2) The instrument's validity is predicated on the zero conditional mean independence assumptions. These conditions imply that, given the instrument, the error terms in the first and second stages of the 2SLS have an expected value of zero, signaling the absence of omitted variable bias and the proper isolation of the instrument's exogenous variation.

(3) The relevance of the instrument is affirmed by a positive correlation between the instrument ($Z_i$), and the endogenous treatment variable ($D_i$). This critical assumption ensures that the instrument exerts a significant influence on the treatment variable, thus excluding the presence of a weak instrument which could undermine the statistical power and the consistency of the estimator.

(4) The technical feasibility of the model estimation is predicated on the variance conditions, specifying that the conditional variances are finite. This technical assumption ensures that the model does not suffer from the problems of infinite variance, which can occur with heteroskedastic errors and lead to unreliable standard errors and test statistics.

**Empirical Results**

Table 2: 6 Regressions Results

*t statistics in parentheses*

* $p < 0.05$, ** $p < 0.01$, *** $p < 0.001$

|  | (1) | (2) | (3) | (4) | (5) | (6) |
|---|---|---|---|---|---|---|
|  | Not Happy Above Median | Not Happy Below Median | Pretty Happy Above Median | Pretty Happy Below Median | Very Happy Above Median | Very Happy Below Median |
| Treatment | -0.203*** | -0.0642** | -0.113*** | 0.173*** | 0.0133 | 0.0662*** |
|  | (-8.67) | (-2.90) | (-4.45) | (6.59) | (1.07) | (4.81) |
| #HH Members | -0.0109* | -0.0294*** | -0.000173 | 0.0252*** | 0.00920** | 0.0142*** |
|  | (-2.17) | (-6.41) | (-0.03) | (4.33) | (3.21) | (4.42) |
| Age | 0.00237*** | 0.00193*** | 0.00151*** | -0.000304 | -0.000210 | -0.000490*** |
|  | (12.62) | (11.14) | (7.00) | (-1.38) | (-1.92) | (-3.94) |
| Female | 0.0264*** | -0.0359*** | 0.0690*** | -0.0284*** | 0.0160*** | -0.00520 |
| | (5.66) | (-7.84) | (13.92) | (-5.34) | (6.66) | (-1.86) |
| Initial Health Condition | 8.05e-12*** | -4.09e-12** | 5.80e-13 | -2.55e-11*** | -2.91e-12*** | -8.19e-12*** |
| | (5.39) | (-2.90) | (0.35) | (-14.82) | (-3.64) | (-8.94) |
| _cons | 0.0624*** | 0.104*** | 0.0777*** | 0.137*** | 0.0187** | 0.0348*** |
| | (4.99) | (8.83) | (5.51) | (9.42) | (2.62) | (4.48) |
| N | 21413 | 21413 | 21413 | 21413 | 21413 | 21413 |

Table 2 shows the result for the six 2sls IV regressions, regression 1 nothappy_above is when $Y_i = 1$ (not too happy) and $〖AM〗_i = 1$ (expenditure above median), regression 2 nothappy_below is when $Y_i = 1$ (not too happy) and $〖AM〗_i = 0$ (expenditure below median); regression 3 pretty_above is when $Y_i = 2$ (pretty happy) and $〖AM〗_i = 1$ (expenditure above median), regression 4 pretty_below is when $Y_i = 2$ (pretty happy) and $〖AM〗_i = 0$ (expenditure below median); regression 5 very_above is when $Y_i = 3$ (very happy) and $〖AM〗_i = 1$ (expenditure above median), regression 6 very_below is when $Y_i = 3$ (very happy) and $〖AM〗_i = 0$ (expenditure below median).

The treatment coefficient $β_1$ in regression (1) is significantly negative at -0.203 (p<0.01), showing that with the Medicaid coverage leads to a substantive decrease in the probability that someone is both in the 'not too happy' category, and has above median spending. On net, the Medicaid coverage is decreasing the number of people that are in the not too happy and above median spending category. This is consistent with the hypothesis that Medicaid coverage makes people happier without worrying about medical expenditures and without increasing their medical expenditures. Contrastingly, for the same happiness category with expenditures below the median (regression (2)), the treatment effect is smaller but still negative and significant at $β_(2) = -0.0642$ (p<0.05), indicating that enrolling in OHP reduces the number of people in this category as well, although by less than for the unhappy and above median category.

For those who were 'very happy', Regression (5) does not show a statistically significant correlation between higher expenditures and happiness $β_(5) = 0.0133$ (p>0.05), suggesting that for this group, the level of spending under the OHP does not have a discernible impact on the fraction of people in this category. Yet, for those with lower expenditures (Regression (6)), a positive and significant treatment

26

effect $\beta_{(6)}= 0.0662$ (p<0.01) is evident, underscoring again that reduced financial strain has increased the probability of people falling into the category of very happy and below median spending post-treatment.

Regression (3) shows that treatment reduces the number of people who are pretty happy and have above-median spending, with a significant and negative coefficient $\beta_{(3)} = -0.113$ (p < 0.01). However, we do not know where people go from this cell: the decrease in probability of people in the pretty happy group post-treatment could have gone to the very happy category or the not too happy category. When expenditures are below the median (regression (4)), the treatment has a positive effect $\beta_{(4)} = 0.173$ (p<0.01), indicating an increase in the number of people under the pretty happy and below median spending category. This result aligns with the hypothesis that reduced financial burden, as facilitated by the OHP, can improve mental well-being.

In summary, enrolling in OHP has significantly decreased the probability of experiencing unhappiness, regardless of whether individuals had high or low medical spending. Additionally, it revealed that, on net, there is a significant decrease in probability of being not too happy and having above median medical expenses post treatment, and a significant increase in probability of being very happy and having below median expenses post treatment. Concerning the overall result, enrolling in the OHP had a positive effect on being happier and spending less. These findings align with the hypothesis that alleviating financial burden as a benefit of Medicaid contributes to the well-being of individuals in distinct subgroups.

**Conclusion and Discussion**

This study's analysis, predicated on a randomized controlled design, delved into the impacts of Medicaid coverage on mental health by observing a low-income, uninsured adult population over approximately one year. The empirical findings reveal an overall trend: a decrease in the number of individuals who are 'not too happy' and have above median spending, and an increase in the number of individuals under the below-median expenditures and 'very happy' category. These findings align with the hypothesis that reduced financial strain after Medicaid contributes to mental health improvement.

The implications of these findings are profound, especially when considering policy designs aimed at optimizing both the health and overall well-being of low-income populations. With the expansion of Medicaid eligibility under the Patient Protection and Affordable Care Act, understanding these effects takes on a heightened policy relevance (Begley et al., 2013). As the OHIE continues to provide a wealth of data, this paper aims to contribute to the dialogue on how Medicaid coverage affects not just health in the narrow sense, but the broader psychosocial well-being of individuals.

For robustness checks, an alternative approach I thought of could have been employed wherein the outcome variable $Y_i$ is substituted with another indicator of happiness, and the regression is re-executed. This would help ascertain the consistency of the observed effects across different measures of well-being.

Future research could explore the causality in this relationship further. Additionally, including interaction terms in the regression model could offer more insights into how the relationship between Medicaid coverage and happiness may be moderated by factors such as health status or financial stress.